\documentclass[twocolumn,showpacs,prl]{revtex4}% Physical Review Letter

\usepackage{graphicx}% Include figure files
\usepackage{amssymb} 
\usepackage{dcolumn}% Align table columns on decimal point
\usepackage{bm}% bold math

\begin{document}

\title{Influence of Chain Interdiffusion Between Immiscible Polymers on Dewetting Dynamics}

\author{S\'everine Copp\'ee$^{1}$}
\author{Sylvain Gabriele$^{1}$}
\author{Alain M. Jonas$^{2}$}
\author{Jacques Jestin$^{3}$}
\author{Pascal Damman$^{1}$}
\email{pascal.damman@umh.ac.be}

\affiliation{
$^1$Laboratoire Interfaces \& Fluides Complexes, Universit\'e de Mons Hainaut, 20, Place du Parc, B-7000 Mons, Belgium
$^2$Unit\'e de Physique et de Chimie des Hauts Polym\`eres, Universit\'e catholique de Louvain, 1, Place Croix du Sud, B-1348 Louvain-la-Neuve, Belgium
$^3$Laboratoire L\'eon Brillouin (CEA-CNRS), CE-Saclay, 91191 Gif-sur-Yvette Cedex, France
}

\date{\today}

\begin{abstract}
The interface between two immiscible polymers, polystyrene (PS) and polydimethylsiloxane (PDMS), was studied by neutron reflectivity and dewetting by using free PS chains and PDMS brushes. Unexpectedly, we found that the PS chains diffuse in the PDMS brushes at temperatures well below the glass transition temperatures of PS, the dynamics being largely determined by the grafting density of the brush. By this study, we demonstrate the major influence of the chains interdiffusion on the friction properties for a couple of immiscible polymers. By the way, the puzzling ageing of PS thin films observed from dewetting experiments is found to be directly related to modifications of the PS/PDMS interface. 
\end{abstract}


\maketitle
%INTRO
%intro Brush
Layers of tethered chains attached by one end to a surface, are particularly important to understand fundamental processes dominated by interfaces such as adhesion, friction or lubrication \cite{Jones99,Leger} and for important technological applications such as stabilization of colloids and patterning of block copolymers.

For a polymer brush in contact with melted chains of the same nature (symmetrical systems), a key parameter is the degree of chain interdiffusion between the brush and the bulk polymer melt. Experimental and theoretical studies showed that the degree of interdiffusion is described by a phase diagram with different regimes \cite{Brown94,Leger}. Depending on the degree of polymerization of brush and free chains and on the grafting density, $\sigma$, a total interdiffusion (small $\sigma$) or a completely dry brush (exclusion of free chains for large $\sigma$) is observed \cite{Vilmin04}. 

To our knowledge, there is however no study about interdiffusion between polymer brush in contact with melted chains of different nature (asymmetric systems), although we expect that the degree of interdiffusion should drastically influence adhesion and friction properties.
% Morphology of the Interface between immiscible polymers
Numerous studies showed that the interface between two bulk immiscible polymer melts is  not atomically sharp, even when the degree of immiscibility is strong such as polystyrene (PS) and polymethylmethacrylate \cite{Sferrazza97}. Finite interfacial width appears either due to partial interdiffusion because the unfavorable enthalpy of mixing is balanced by a gain in chain
entropy \cite{Bucknall04,Geoghegan}, or due to capillary fluctuations of the vertical position of the interface \cite{Sferrazza97}. Considering this situation, we could expect that interdiffusion should also occur for asymmetric brush/melt systems. However, the extent of this interdiffusion, the influence of grafting density, the existence of various regimes and the influence of the degree of interdiffusion on friction properties remains to be established. 

In this paper, we will report a study of the interface between PS free chains in contact with polydimethylsiloxane (PDMS) brush, a fully asymmetric system. Neutron and x-ray reflectometry was used to probe the structure of the burried PS/PDMS interface while dewetting dynamics gave informations about the global friction of free chains at the interface.
%Why dewetting
Dewetting is indeed highly sensitive to minute modifications of the interface provided to be in a slipping regime. The importance of the interface on dewetting dynamics can be understood by considering that long polymers on nonadsorbing substrates slip at the fluid/substrate interface \cite{degennes85, Leger, Brochard98}. 
The degree of slipping is usually caracterized by the slippage length, $b$ (obtained from the extrapolation to zero velocity) \cite{degennes85}. 
Autophobic dewetting of PDMS thin films (free chains) on various PDMS brushes, a symmetric geometry, have recently highlighted the major role of the interface on friction dynamics for chain-like molecules \cite{Reiter01a,Casoli01}. 

Understanding interdiffusion between immiscible polymers will also shed new light on the properties of PS thin films obtained by spin-coating, a topic of intense research for years. Very recently, we have highlighted the major role of the out of equilibrium chain conformations and the lack of entanglements, resulting from fast solvent evaporation, on the viscoelastic properties of these spin-coated thin films \cite{Damman07}. However, a puzzling observation still remains unsolved. Surprisingly, the age of these PS thin films has an amazing influence on the dewetting dynamics, even for ageing temperatures deeply below the glass transition temperature of bulk PS (e.g., $80^{\circ}$C below $T_g$) \cite{Reiter05}. 
Based on a detailed study of the PS/PDMS interface, we will also give an answer to this puzzling ageing effect.

%%EXPERIMENTAL SECTION
The PDMS monolayers were obtained by annealing spin-coated films at various adsorption temperatures (from 80$^{\circ}$C to 200$^{\circ}$C) for 1 hour \textit{in vacuo} ($M_w = 90.2 \ kDa$, polydispersity index, $I_p$ = 1.96). The non adsorbed PDMS chains were removed by immersing the silicon wafer in heptan. PS thin films ($M_w = 120 \ kDa$, $I_p=1.04$) were spin-coated from toluene solutions onto silicon substrates previously decorated with a PDMS brush. The thickness of the PDMS and PS films were measured by ellipsometry. Ageing was performed by maintaining the samples at 80$^{\circ}$C (for PS, $T_g=100^{\circ}C$) for various periods of time. Isothermal dewetting at 130$^{\circ}$C of thin PS films was followed by optical microscopy. The dewetted distance $D$ and the rim width $W$ were measured by interference image contrast. 
%NEUTRON REFLECTIVITY
For the reflectivity experiments, we used Si/h-PDMS/d-PS systems giving the best contrast. Samples measurements were performed on the time-of-flight reflectometer EROS, on the Orph\'ee reactor at the Laboratoire L\'eon Brillouin (CEA, Saclay). The time-of-flight technique allows us to irradiate the sample with a range of wavelengths at the same time, minimising the required number of angles. In our case, data were obtained at two angles, 1.36$^{\circ}$ and 2.5$^{\circ}$, with wavelengths in the range 0.3 nm  to 3.0 nm allowing an extremely good range in neutron wavevector. The neutrons beam enters through the silicon substrate (5 mm thick). The theoretical scattering length density values used throughout the study are : Si = 2.07; SiO$_2$ = 3.4; {\it h}-PDMS = 0.0639; {\it d}-PS = 6.0 (in $10^{-6} \textrm{\AA}^{-2}$).
%X-RAY REFLECTIVITY
X-Rays reflectivity measurements were performed on the very same samples to characterize the PS/air interface. The goniometer is a modified Siemens D5000 2-circles fitted to a Rigaku rotating anode delivering Cu K($\alpha$) radiation of a 0.15418 nm wavelength.
%Fitting
The neutron and X-ray data were analysed using standard fitting procedures \cite{Arys01}, using a single set of geometrical parameters for both x-ray and neutron reflectometry data. 

\begin{figure} 
\includegraphics[width=5cm]{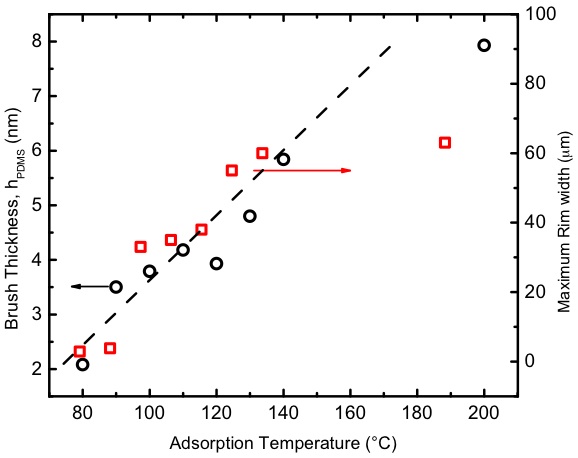}
\includegraphics[width=5cm]{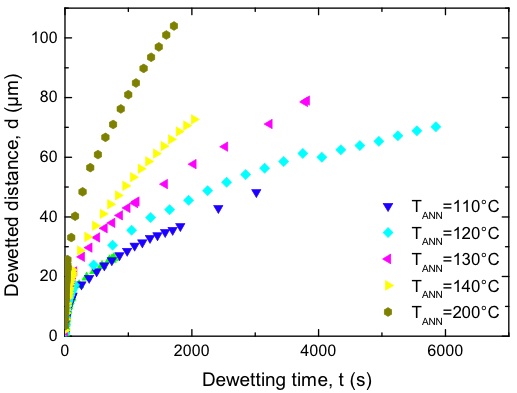} 
\caption{(a) Thickness of PDMS monolayers irreversibly adsorbed on silicon wafer at various temperatures, $T_{ads}$ together with the evolution of the maximum rim width recorded during dewetting of PS on PDMS. (b) Dewetting dynamics (dewetted distance, $D$, vs time) for PS thin films spincoated onto PDMS monolayers (thickness, $h_{PS} \sim 100$ nm, dewetting temperature, $T_d = 130^{\circ}$C).} 
\label{PDMS}
\end{figure}

%%RESULTS and discussion
The adsorption of hydroxyl terminated PDMS chains on silicon wafers corresponds to a reversible adsorption of end-tethered polymer chains, an intermediate situation with respect to polymer brush (grafted chains on repulsive surfaces) and reversibly adsorbed monolayers. As shown previously, these monolayers adopt a brush-like morphology since the reversible adsorption only affects a marginal region close to the surface, its extent being determined by the grafting density, $\sigma$ \cite{Raphael96}. 
Due to a progressive grafting of the chains (a thermally activated process), the thickness of the irreversibly adsorbed PDMS monolayer increases with the adsorption temperature indicating a progressive stretching of the PDMS chains to finally form a dense brush (Figure \ref{PDMS}). The reduced grafting density is given by $\Sigma = \sigma \pi R_g^2$, where $R_g$ is the radius of gyration ($R_g = a \sqrt{\frac{N}{6}} \sim 8$ nm, $a$ being the length of a Kuhn segment). Considering a simple melt brush model \cite{Milner91}, the grafting density can be deduced from the monomer density, $\rho$, equal to $\rho = N \sigma / h$ (where $N$ and $h$ are the number of Kuhn segments in a chain and the brush thickness, respectively). For the lowest adsorption temperatures, reversible adsorption dominates, $\Sigma$ being close to 2.5. The PDMS monolayers adopt a ''loops and tails'' morphology \cite{Raphael96,Leger}. In contrast, high temperatures ($T_{ads} = 200 ^{\circ} C$) enhance the grafting of PDMS chains (via the hydroxyl end groups) and yield $\Sigma$ values as high as 11. The monolayers adopt a brush-like morphology. 

%Dewetting dynamic
With highly immiscible polymers such as PS and PDMS, the lack of chains interpenetration at the interface would naively suggest that modification of the brush morphology will exert a very limited influence on the burried interface and thus on the dewetting dynamics. However, as shown in Figure \ref{PDMS}, the thickness of the PDMS brush drastically affects the dewetting dynamics of PS films. More surprisingly, the dynamics is accelerating when the brush thickness increases. At first sight, the reverse evolution was expected since thicker PDMS brushes will dissipate more energy by the deformation of the adsorbed PDMS chains leading thus to slower dewetting velocities. This apparent contradictory evolution will be explained later in the light of the neutron reflectivity results. 
As shown previously, the dewetting dynamics of PS thin films on coated silicon wafers is dominated by the friction of chains at the interface and the viscoelasticity of PS \cite{Brochard98,Vilmin06,Damman07}. In that regime, a simple balance of capillary and dissipation energy gives the dewetting velocity, $V = \frac{S}{\eta} b W^{-1}$ where $S$, $\eta$, $b$ and $W$ are the spreading coefficient, the melt viscosity, the slippage length and the rim width, respectively\cite{Vilmin06}. 
The size of the rim collecting the dewetted fluid is determined by the propagation of the elastic stress in the film which in turn  is determined by  the reptation time (or the viscosity, $\eta$) and the friction of the polymer chains on the substrate (i.e., friction coefficient, $k$). The rim width can be estimated by the relation  $W \approx \Delta = \sqrt{b h_{PS}}$ (where $h_{PS}$ is the film thickness, the slippage length $b = \eta/k$) \cite{Brochard98}. Variation of rim width is thus a clear indication of a change in the slippage length. 
From the evolution of the maximum rim width (Figure \ref{PDMS}(a)), we could infer that the slippage length increases by more than two orders of magnitude by changing the morphology of the PDMS layer ($b \propto W^2$). Even for a couple of immiscible polymers, tiny changes at the polymer/polymer interface induce severe modifications of the friction properties and consequently of dewetting dynamics. 

\begin{figure} 
\includegraphics[width=6cm]{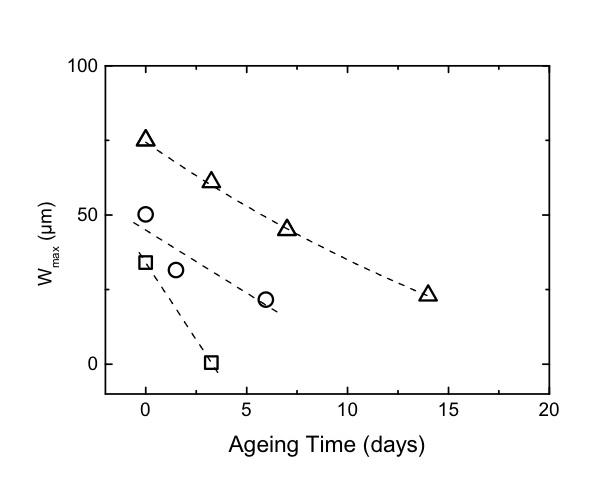}
\caption{Influence of ageing time on the dewetting dynamics illustrated by the plot of maximum rim width, $W_{max}$, with ageing/annealing time for different adsorption temperatures of the PDMS chains ($T_{ads} =$ 100 ($\square$), 140 ($\bigcirc$) and 200 $^{\circ}C$ ($\bigtriangleup$).}
\label{ageing}
\end{figure}

%Ageing/annealing
These observations open new perspectives to explain the puzzling ageing of PS thin films as deduced from dewetting experiments \cite{Reiter05}. 
As shown in Figure \ref{ageing}, this ageing phenomenon is also observed for the dewetting of PS films on the various PDMS monolayers used in this study. However, as shown by the evolution of the maximum rim width, a clear indicator of the dewetting dynamics (Figure \ref{ageing}), ageing is much more pronounced for the thinnest layers (''loops and tails'' morphology, $T_{ads} \le 100^{\circ}$C). Indeed, PS films become apparently stable on these PDMS monolayers after 2 days of annealing at a temperature well below $T_g$. Very much larger annealing times are required to observe significant decrease of the rim width for the PDMS brushes prepared at high adsorption temperatures ($T_{ads} \sim 200^{\circ}$C). The thickness of the PDMS monolayers thus affects unexpectedly the dynamics of ageing.

As shown here above the slippage length of PS chains is directly related to the maximum rim width ($W \sim \sqrt{b h_{PS}}$) and strongly depends on the morphology of the interface. Based on these considerations, we suggest that the slowing down of the dewetting dynamics observed during ageing is directly related to a relaxation of the PS/PDMS interface, i.e., an increase of the chain interdiffusion. 
In contrast to relaxation processes in bulk PS, a reorganization of the PS/PDMS interface can be observed at very low temperatures. Interdiffusion of PS and PDMS chains could lead to a thin interfacial layer of higher chain mobility. The interfacial glass transition temperature, $T_g^i$ is given by $1/T_g^i = \phi_{PDMS}/T_g^{PDMS}+\phi_{PS}/T_g^{PS} \sim (214 K)^{-1}$ with equivalent fraction of both polymers ($\phi_{PDMS} = \phi_{PS} = 0.5$) \cite{Rubinstein}. 

\begin{figure}
\includegraphics[width=6cm]{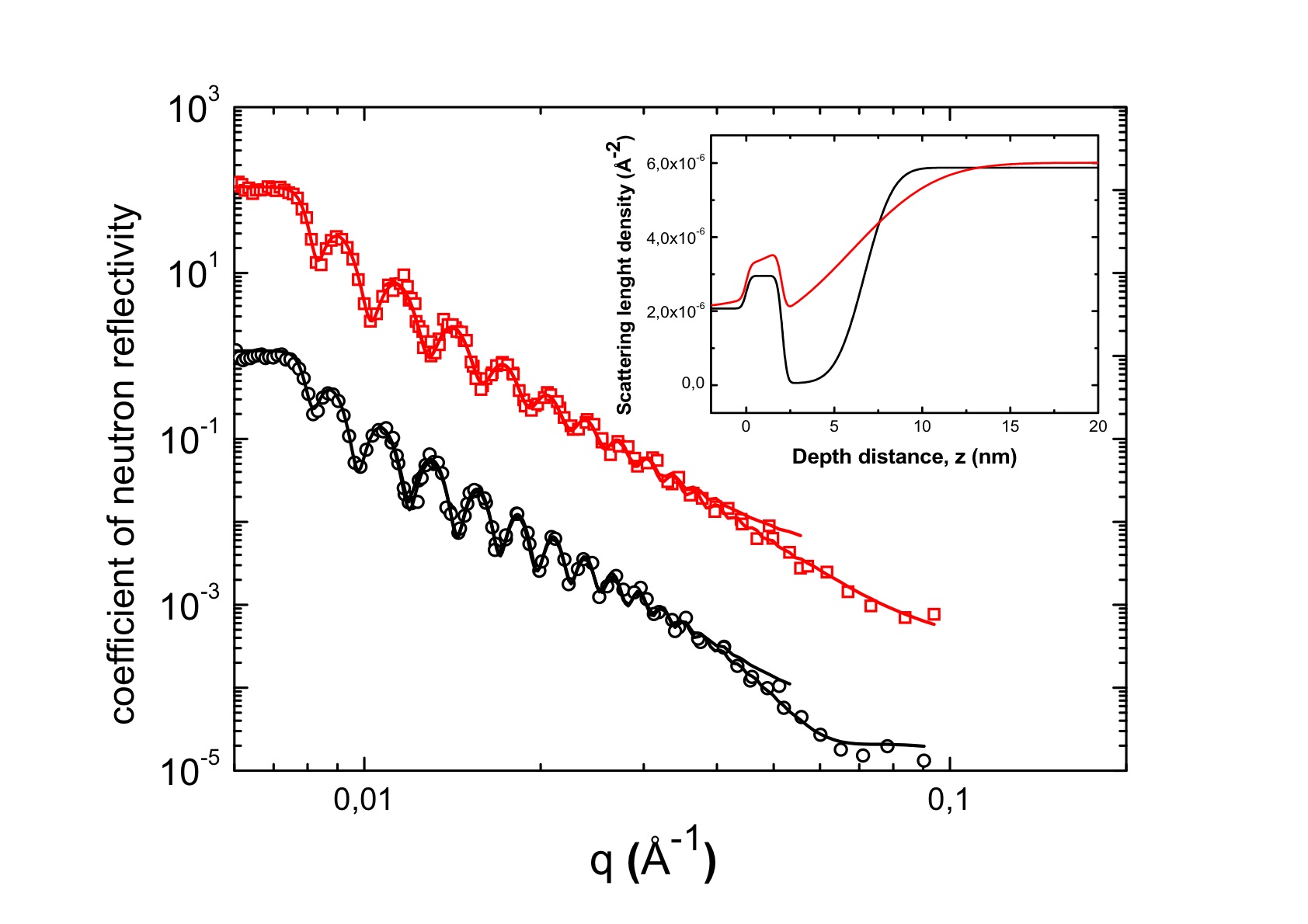}
\caption{Neutron reflectivity curves of \emph{Si/SiOx/h-PDMS/d-PS} multilayers ($T_{ads} = 100 ^{\circ}$C) for fresh ($\bigcirc$) and aged ($\square$) samples (ageing time of 68 h at 80 $^{\circ}$C). The inset shows the scattering length density depth profiles resulting from the fitting of reflectivity curves.}
\label{neutron}
\end{figure}

%Reflectivity - interfacial width
To check the validity of this new hypothesis, we have directly probed the PS/PDMS interface by combined neutron and X-ray reflectivity for fresh and aged samples (ageing at $80^{\circ}C$ for 68 h). 
%Neutron reflectivity is a very powerful method of analysing buried interfaces. 
The large difference in neutron scattering length density between deuterium and hydrogen makes possible experiments using isotopic substitution that induces a nice contrast at the interface between two polymer thin films. 
The neutron and X-ray reflectivity data were fitted to obtain the evolution of the scattering density with depth, $z$, (Figure \ref{neutron}). These $z$ profiles directly show the layer thicknesses and, more importantly, the root-mean-square (rms) width of the interface between PS and PDMS, $\sigma_i$. We should take care however that the observed interfacial width (or roughness) can be related either to the interdiffusion of deuterated and hydrogenated chains or to local fluctuations of the vertical position of the interface. However, due to the grafting, the lateral mobility of PDMS chains is suppressed and the monolayer thicknesses are very small (less than 10nm). We thus assume that the observed rms interfacial width is mainly related to the degree of interdiffusion of PS and PDMS chains.

\begin{figure}
\includegraphics[width=6cm]{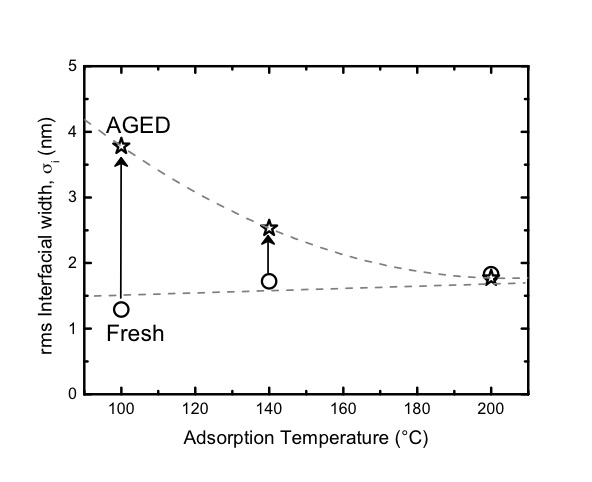}
\caption{Evolution of the PS/PDMS rms interfacial width, $\sigma_i$, with the PDMS adsorption temperature, $T_{ads}$, for fresh and aged samples (ageing/annealing times of 68 h at 80 $^{\circ}$C). }
\label{interface}
\end{figure}

Evolutions of the interfacial width, $\sigma_i$ with the PDMS adsorption temperature are shown in Figure \ref{interface} for fresh and aged samples. For fresh samples, the scattering density vs $z$ profiles show a almost constant rms interfacial width ($\sigma_i \sim 1.3 - 1.8 nm$) with little influence of the monolayer morphology, brush or ''loops and tails'' (Figure \ref{interface}). In contrast, the ageing of the interfacial width depends significantly on the grafting density of the PDMS monolayer. While no change is observed for high adsorption temperatures (high grafting density, true brush morphology), the rms interfacial width increases up to $4 nm$ for the ''loops and tails'' monolayers prepared at $100^{\circ}C$. 
The reflectivity curves show without any doubt that the PS and PDMS chains progressively interdiffuse at the interface during ageing even for temperatures well below the $T_g$ of PS. We also found that the interdiffusion dynamics strongly decreases when the grafting density of the PDMS chains increases (Figure \ref{interface}). For very high grafting densities, free PS chains can be completely excluded from polymer brushes \cite{degennes80,Milner91,Vilmin04}. 
At the end of ageing, the interdiffusion is so important for the lowest grafting density that a pure PDMS layer can no more be observed in the $z$ profile of scattering density (see inset of Figure \ref{neutron}). 

The structural changes of the burried PS/PDMS interface during ageing explain the dewetting experiments. Sliding friction experiments showed that chain interdiffusion at the interface drastically affects the friction properties of polymers \cite{Leger,Casoli01}. The degree of slippage of PS chains on the PDMS brushes, directly determined by the friction ($b = \eta/k$), will thus decrease as the chains interdiffuse at the interface. This decrease of slippage length, $b$, induces a drastic decrease of the maximum rim width ($W=\sqrt{bh_{PS}}$, Figure \ref{ageing}) and a slowing down of the dewetting velocity as observed. 
The influence of ageing on the dewetting dynamics can thus finally be related to a gradual diffusion of PS chains in the PDMS brushes. Thanks to the very high mobility of PDMS chains, this diffusion occurs well below the bulk glass transition temperature of PS.

%CONCLUSIONS - Dewetting + reflectivity, unified picture !
From neutron reflectivity and thin film dewetting, the puzzling ageing of PS thin films was found to be related to a modification of the PS/PDMS interface. Interestingly, this study also highlights the major influence of the chains interdiffusion for friction and slippage even for immiscible polymers.
This study demontrates again that dewetting is particularly sensitive to minute changes of the buried interface and thus confirms its ability to probe the nature of the interfaces between two immiscible polymers.

%ACKNOWLEDGMENT
\acknowledgments
The authors gratefully acknowledge E. Rapha\"el, F. Ziebert and G. Reiter for many fruitful discussions and suggestions.
This work was supported by the Walloon Region (CORRONET Research project) and the Belgian National Fund for Scientific research (FRS-FNRS). S\'everine Copp\'ee thanks the ''Fonds pour la Formation \`a la Recherche dans l'Industrie et dans l'Agriculture'' (FRIA) for financial support. The authors thank M. Vou\'e and N. Dahmouchene for ellipsometry measurements and A. Moussa and Z. Hu for X-ray reflectivity measurements. S. Gabriele is an associate of the FNRS.

\newpage

\end{document}